%% file: main.tex
\documentclass[a4paper,conference]{IEEEtran}
\IEEEoverridecommandlockouts
\usepackage{cite}
\usepackage{amsmath,amssymb,amsfonts}
\usepackage{mathtools}
\usepackage{bm}
\usepackage{algorithmic}
\usepackage{graphicx}
\usepackage{textcomp}
\usepackage{xcolor}
\usepackage[binary-units=true]{siunitx}
\usepackage{booktabs}

\usepackage[]{footmisc}
\usepackage[hidelinks]{hyperref}
\usepackage{caption}
\usepackage{subfig}

\usepackage{listings}
\usepackage{xcolor}
\usepackage{setspace}

\definecolor{codegreen}{rgb}{0,0.52,0.24}
\definecolor{codegray}{rgb}{0.5,0.5,0.5}
\definecolor{codepurple}{rgb}{0.62,0.078,0.59}
\definecolor{backcolour}{rgb}{0.94,0.94,0.94}

\lstdefinestyle{mystyle}{
  backgroundcolor=\color{backcolour},   commentstyle=\color{codegreen},
  keywordstyle=\color{codepurple},
  numberstyle=\tiny\color{codegray},
  stringstyle=\color{codepurple},
  basicstyle=\ttfamily\footnotesize,
  breakatwhitespace=false,         
  breaklines=true,                 
  captionpos=b,                    
  keepspaces=true,                 
  numbers=left,                    
  numbersep=2pt,                  
  showspaces=false,                
  showstringspaces=false,
  showtabs=true,                  
  tabsize=2
}

\newcommand{\slice}{\texttt{SL}}
\newcommand{\slices}{\texttt{SLs}}
\newcommand{\sne}{\texttt{SNE}}
\newcommand{\xbar}{\texttt{C-XBAR}}
\newcommand{\streamer}{\texttt{DMA}}
\newcommand{\streamers}{\texttt{DMAs}}
\newcommand{\collector}{\texttt{collector}}
\newcommand{\cluster}{\texttt{Cluster}}
\newcommand{\clusters}{\texttt{Clusters}}
\newcommand{\sequencer}{\texttt{Sequencer}}

\newcommand{\arpan}[1]{{\color{black}#1}}

\def\BibTeX{{\rm B\kern-.05em{\sc i\kern-.025em b}\kern-.08em
    T\kern-.1667em\lower.7ex\hbox{E}\kern-.125emX}}
\begin{document}

\title{SNE: an Energy-Proportional Digital Accelerator for Sparse Event-Based Convolutions\\
{}
}

\author{\IEEEauthorblockN{Alfio Di Mauro\IEEEauthorrefmark{2}, Arpan Suravi Prasad\IEEEauthorrefmark{2}, Zhikai Huang\IEEEauthorrefmark{2}, Matteo Spallanzani\IEEEauthorrefmark{2},  Francesco Conti\IEEEauthorrefmark{3}, Luca Benini\IEEEauthorrefmark{2}\IEEEauthorrefmark{3}}
\IEEEauthorblockA{\IEEEauthorrefmark{2}Dept. of Information Technology and Electrical Engineering, ETH Z\"{u}rich, Switzerland} \IEEEauthorblockA{\IEEEauthorrefmark{3}Dept. of Electrical, Electronic and Information Engineering, University of Bologna, Italy}}


\maketitle

\begin{abstract}
Event-based sensors are drawing increasing attention due to their high temporal resolution, low power consumption, and low bandwidth. To efficiently extract semantically meaningful information from sparse data streams produced by such sensors, we present a 4.5TOP/s/W digital accelerator capable of performing 4-bits-quantized event-based convolutional neural networks (eCNN). Compared to standard convolutional engines, our accelerator performs a number of operations proportional to the number of events contained into the input data stream, \textcolor{black}{ultimately achieving a high energy-to-information processing proportionality}. On the IBM-DVS-Gesture dataset, we report 80uJ/inf to 261uJ/inf, respectively, when the input activity is 1.2\% and 4.9\%. Our accelerator consumes 0.221pJ/SOP, \arpan{to the best of our knowledge} it is the lowest energy/OP reported on a digital neuromorphic engine.
\end{abstract}


\setstretch{0.97}

\input{content/introduction}
\input{content/related_works}

\input{content/architecture}

\input{content/simulation_results}

\input{content/conclusion}

\section{Acknowledgement}
This work has been supported in part by the \textit{Ampere} project funded by the EU Horizon 2020 research and innovation programme under grant agreement No. 871669.
\textcolor{black}{The authors would like to thank \textit{Armasuisse Science and Technology} for funding this research, and \textit{IniVation} for kindly lending us a DVS camera.  }

\bibliographystyle{IEEEtran}
{\bibliography{bibliography}}

\end{document}

%% file: content/introduction.tex
\section{Introduction}

In recent years, we experienced a profound transformation in how digital systems deployed around us operate. Edge-computing devices evolved toward lower power consumption, longer lifetime, heterogeneity, and higher computational capabilities. Similarly, sensors improved to efficiently convey information to processing engines while balancing low power consumption, responsiveness, and energy spent on data transfer.

Event-based sensors are an emerging class of devices that measure a physical quantity and transfer such information in a frame-less fashion.
Event-based vision sensors (EVSs), which asynchronously measure the brightness changes in the field of view, belong to this category. Compared to a traditional frame-based imaging sensor, an EVS outputs a stream of events that encodes the brightness change's time, location, and polarity. EVSs typically feature higher temporal resolution (in the order of few \SI{}{\micro \second}) and a lower power consumption (ranging from \SI{250}{\micro \watt} to \SI{2}{\milli \watt}); in low activity scenarios, their bandwidth can be as low as a few kB/s\cite{dvsInivation}.

A key advantage introduced by event sensors is the proportionality between \textcolor{black}{the primary sensor input and the number of output events generated by it \cite{FlyDVS2021}. To efficiently exploit the inherently sparse nature of such data streams, the energy to information proportionality needs to be preserved across the whole processing pipeline. 
CPU and GPU class devices can marginally profit from unstructured data sparsity, while dedicated deep neural networks (DNN) accelerators often rely on dedicated architectural features tailored to the specific type of data sparsity to achieve high energy efficiencies in such scenarios \cite{Baoxin2020}. To overcome this limitation, a paradigm shift in how such sparse data are processed is needed.}

Neuromorphic algorithms, i.e. algorithms inspired by how biological brains work, are promising candidates to solve the unstructured data processing problem. Among neuromorphic algorithms, Spiking Neural Networks (SNNs) represent the leading model-free algorithmic approach for EVSs data processing \cite{Roy2021}. Like many artificial neural networks (ANNs), SNNs rely on elementary computational units, i.e. neurons, which are interconnected through synaptic weights to form computational networks\cite{Panda2020}. A distinctive feature of SNN neurons is the presence of a \textit{neuron internal state}, which evolves over the entire inference process.
Recent advances in SNNs show that such a class of networks can achieve accuracy levels comparable to state-of-the-art (SoA) deep learning networks while significantly reducing the number of required computational operations \cite{Sengupta2019a}, therefore making them a suitable candidate to achieve high energy-to-information processing proportionality. 

In this work, we present a novel digital sparse neural engine (SNE) to efficiently accelerate SNN inference tasks at the extreme edge. Our accelerator exploits an explicit input event temporal and spatial location encoding, the \sne~architecture is designed to improve input data and weight reuse, reducing the traffic towards the memory. \sne~achieves a maximum performance of 51.2 GSOP/s, and an energy efficiency of 4.5TSOP/s/W. Ultimately, \sne~shows 3.55X higher energy efficiency than SoA neuromorphic platform \cite{Pei2019}, approaching classical DNN accelerators energy efficiencies \cite{classical_dnn}, while performing energy-proportional computations. As a proof of concept, we show that \sne~consumes 0.221 pJ/SOP and achieves 92.8\% accuracy on a classification task performed on the IBM DVS-Gesture data set.

%% file: content/related_works.tex
\section{Related works}
Over the last years, research and industry have proposed various deep learning engines to accelerate inference at the edge, achieving extreme energy efficiencies \cite{reuther}. As algorithmic research proposed low-precision, highly-quantized networks, hardware platform evolution kept the pace by proposing new architectures capable of exploiting low memory footprints and performing low-precision operations\cite{sebe,dimauro2020}.
Recently, neuromorphic algorithms have been attracting increasing attention as a  more energy-efficient alternative to conventional deep learning approaches \cite{beigne2019}, especially in those contexts where input feature maps are produced at a nonconstant rate and are also characterized by high unstructured sparsity\cite{Young2019a}. 

Such algorithms run on neuromorphic platforms that can be divided into two main categories: Analog and mixed-signal, and digital SNN accelerators. Analog and mixed-signal implementation present several advantages over digital implementations, e.g. they typically achieve higher energy efficiencies and smaller neuron area footprint \cite{rubino2020ultralowpower}. Also, they typically implement more complex neuron models. However, these designs are hard to scale, as their functionality is often technology-dependent, requiring laborious tuning to the technology node. Additionally, mixed-signal operations require many biases generated on-chip, often degrading the system-level energy efficiency. Contrarily, digital implementation typically features a less complex and more scalable neuron model\cite{loihi,truenorth,Frenkel2020A2C}, as well as fast integration in digital SoCs and technology porting.

Compared to the accelerators mentioned above, \sne~explicitly encodes the temporal and spatial location of the events to reduce the temporary data memory footprint of highly sparse input and intermediate feature maps. \sne~also maximizes input data and weight reuse, eventually reducing the traffic towards the memory. Compared to the existing neuromorphic platforms (table \ref{tab:comparison}), \sne~performs synchronous parallel execution. This feature, coupled with the explicit event encoding, compresses long intervals of sparse input activity into dense computational phases performed at high frequency. Our accelerator improves the SoA in energy efficiency by 3.55X while achieving SoA accuracy of 92.8\% on the IBM DVS-Gesture data set.


%% file: content/architecture.tex
\section{Architecture}
\label{sec:architecture}

\subsection{Spiking Neural Network}

A Spiking Neural Network (SNN) is an Artificial Neural Network (ANN) subcategory whose elementary unit mimics biological neurons \cite{Farabet2012a}. Compared to ANNs, SNNs neurons,  named as \textit{spiking neurons}, feature an internal state variable, a \textit{membrane potential}. A firing behavior characterizes spiking neurons, i.e. a spiking neuron generates an impulse on its output when the membrane potential value exceeds a threshold. 
In SNNs, input and output feature maps can be encoded as binary tensors;  the presence of a non-zero value represents the spatial and temporal position of the spike produced by a neuron. Connections among spiking neurons are weighted by synaptic weights. In complex neuron models, synapses can also time-shift input spikes. Synaptic connections between successive layers of an SNN can follow a convolutional or fully-connected scheme.

\subsection{SNE neuron model}
Elementary SNN neurons can be modeled at a different biological plausibility level, ranging from highly approximated neuronal behavior to bio-plausible models\cite{Izhikevich}. In \sne, we implemented a leaky integrate and fire (LIF) neuron. We linearly approximated the exponential membrane potential decay to simplify the hardware design as an iterative linear decay.
The elementary neuron membrane potential update is given by $ V_{mem}[t+1] = -L + \sum_{j}W_{ij}S_{i}[t] $; in our implementation, $L$ is a re-programmable leakage quantity that is subtracted at every time step.
The firing rule is described by $S[t] = \Theta(V_{mem}[t] - V_{th})$, where  $\Theta$ is the Heaviside step function, and $V_{th}$ is a programmable firing threshold.

\subsection{SNE execution model}
\label{sec:exec_model}
The \sne~accelerator data path has been optimized for event-driven computation.
Compared to a standard CNN layer, in event-based convolutional neural network (eCNN) layers we find an additional time dimension. The event-based convolution is performed for each time step, and the state of each neuron resets at the beginning of a new inference. The input synaptic contributions are accumulated in the state variable across the entire inference process.


\begin{figure}[b]
    \centering
    \centerline{\includegraphics[width=0.8\linewidth]{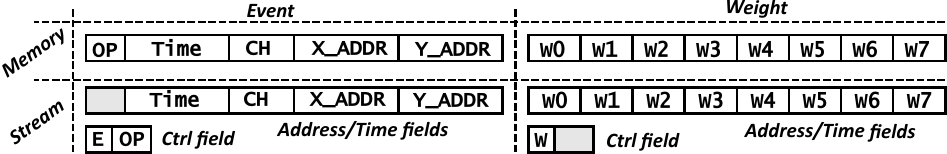}}
    \caption{SNE data format for event and weight.}
    \label{fig:data}
\end{figure}

To exploit the input feature map sparsity, events are encoded explicitly and stored into the memory with the format reported in \figurename \ref{fig:data}; \sne~is fed with individual events instead of input tensor tiles.
Listing \ref{lst:sne_conv} reports the SNE input event processing pseudo-code.
To better exploit spatial and temporal data reuse opportunities,
as an outermost loop, we span the time dimension and process all events occurring at a certain time step. Then, \sne~updates all the output neurons depending on the current input, without spanning the horizontal and vertical dimension of the input tensor in the innermost loop. This loop organization has been chosen because data-path instances are stateful. Therefore, output spike sequences related to each output neuron have to be produced entirely once the input is presented.
Multiple input channels can be accumulated on the same output neuron, \sne~can store up to 256 sets of weights in a filter buffer, and they can be independently selected on-the-fly by each \cluster, according to the addressing of the input event.
An \sne~event is defined by a 32bits value partitioned into the quadruple: $E_i := (OP_e,t,x,y)$. $OP_e$ stands for event operation, and it can be of three different types:
\begin{itemize}
    \item \texttt{RST\_OP} is the event operation that resets the state variable of the neurons to zero.
    \item \texttt{UPDATE\_OP} is the operation that updates the membrane potential of the neurons having the current input event in their receptive field. 
    \item \texttt{FIRE\_OP} is the operation that concludes the neuron state update phase and allows the neuron whose status variable value is above the threshold to produce an output event.
\end{itemize}

\begin{lstlisting}[basicstyle=\ttfamily\footnotesize,float=t,language=Python,label=lst:sne_conv, caption=SNE sparse eCNN layer execution.]
# SW managed loops ---------------------------------
for k_o in range(0,C_o):           #output Ch events
 program_sne(W)                      #change weights 
#--------------------------------------------------- 
  # SNE managed loops ------------------------------
  for t in range(0,T):               #time dimension
   for evt_i in events_in[t]      #explicit evt repr
    k_i,e_x,e_y = get_address(evt_i)
    for i in range(0,H_o):         #output h neurons
     for j in range(0,W_o):        #output v neurons
      w_ij = weight(i,j,k_i,e_x,e_y,W)  #weight calc
      evt_o = neuron_dynamics(i,j,w_ij)
      events_out[t].append(evt_o)      #push evt out
  #-------------------------------------------------
\end{lstlisting}

\subsection{SNE architecture}
\begin{figure}[b]
    \centering
    \centerline{\includegraphics[width=0.9\linewidth]{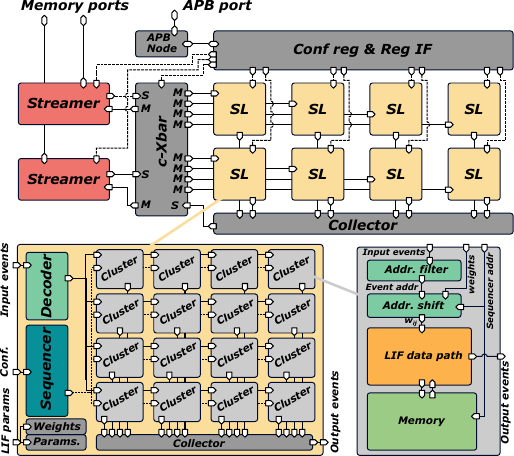}}
    \caption{SNE architecture block diagram.}
    \label{fig:archi}
\end{figure}

\begin{figure}
    \centering
    \centerline{\includegraphics[width=0.95\linewidth]{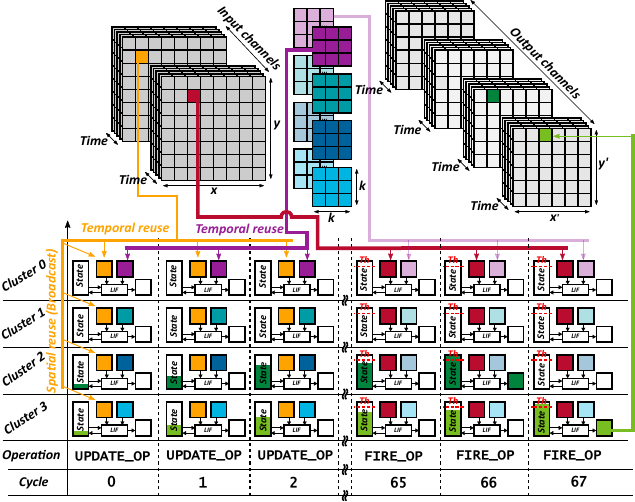}}
    \caption{SNE convolution operation execution.}
    \label{fig:sample}
\end{figure}

The \sne~architecture (\figurename \ref{fig:archi}) is composed by a set of independent parallel processing engines called \arpan{slices (\slices)}. 
Each \slice~is connected to a synaptic crossbar (\xbar), which also connects two autonomous direct memory access engines (\streamer) used to transfer events from the memory to the \slices~and vice versa.
Output event streams produced by the \slices~are joined in a single stream using a \collector, which is also connected as a master to the \xbar.
\sne~can be integrated as a memory-mapped peripheral into a system on chips (SoC) and programmed through a register interface.
The following subsections provide a more detailed description of each \sne~top-level module. 

\subsubsection{c-xbar}
The \xbar~routes both streams of events and weights from the main memory to the slices or vice versa. The data format used for the internal event representation is described in \figurename \ref{fig:data}.
Each \slice~is connected to the \xbar~with communication protocol using a ready-valid (RV) handshake for flow control. The \xbar~can operate in two distinct modes: \textit{i)} single master to single slave port (point-to-point); this configuration is also used to both transfer events and load configuration parameters. \textit{ii)} single master to multiple slave ports (broadcast); in this configuration, the \xbar~can perform flow control and pause the transaction until all slave ports have received the event. 

\subsubsection{streamer}
\streamers~autonomously transfer events and weights from the main memory to the \sne~internal buffers and vice versa. \sne~input events can be stored linearly into the external memory. Therefore, \streamers~implement a simple 1D data movement scheme; they also operate the conversion between the event memory format and event stream format shown in \figurename \ref{fig:data}.
The \streamer~contains a 16-words First-In-First-Out (FIFO) event memory to absorb memory latency cycles (e.g., due to access contention).

\subsubsection{collector}
The \collector~allows packing the output event streams from each \slice~into a single \arpan{time synchronized} stream and sending it to the \xbar. Since the activity of the \slices~is sparse, a single \streamer~can provide significantly more bandwidth than required on a single \slice~output port. Therefore, the collector arbitrates between the \slices~output ports and multiplexes them into a single event stream towards the memory. 

\subsubsection{slices and \arpan{clusters}}
The number of \slices~in \sne~is parametric; the architecture of a single \slice~is shown in \figurename \ref{fig:archi}. Each \slice~instantiates 16 parallel computational units, called \clusters.
Each \cluster~data-path is designed to compute a neuron state update in a single clock cycle, using a single data-path: the implementation of multiple neurons is achieved by time-domain multiplexing (TDM), storing the neuron states in a local latch-based buffer.
Each \cluster~implements 64 TDM neurons using 4 bits for synaptic weights and 8 bits for the internal state. 
Before dispatching input events to the \clusters, the \slice~needs to decode the event operation to perform. 
Units that do not have to update their internal state (i.e., membrane potential) are clock-gated to reduce power consumption. In the case of a \texttt{RST\_OP}, all the \clusters~are activated, and the membrane potential resets for all the neurons of the \slice. All \clusters~of an \slice~receive the same input event. We implemented an address filtering mechanism to selectively redirect input events to specific neurons. 
Execution on all \clusters~happens synchronously and is orchestrated by a module called \sequencer. The \sequencer~provides the address of the current TDM neuron update. 
When the slice is executing a \texttt{SPIKE\_OP}, all the 64 TDM neurons of each \cluster~can potentially produce an output event. To avoid stalling the TDM neurons update, each \cluster~is connected to an output event FIFO, and all FIFOs are connected to a \collector~module.
All computed output neurons across the \clusters~have the same relative position. The absolute spatial mapping of the output neurons is achieved by shifting each address with respect to the \cluster~base address. To achieve high throughput, the following measures were put in place at the \cluster~level:
\textit{i)} the LIF neuron dynamic data path is combinational.
\textit{ii)} To overcome the high memory bandwidth that characterizes SNN inference, the neuron states of each \cluster~are stored locally into two dedicated state memories. During the neuron state update, values are fetched alternatively from each memory and stored the cycle after in a double buffering fashion, practically achieving one state update per cycle.
\textit{iii)} a time-of-last-update (TLU) is stored per \cluster, the next neuron state is computed based on the current timestep value and TLU, skipping the state update in the absence of input activity between two successive timesteps.
\subsubsection{mapping}
The \sne~can be used in two modes. If the neurons of an SNN can be mapped entirely on the \sne~available \clusters~along the spatial dimensions, each \slice~can be used to implement a different layer of the network, and the synaptic connections between neurons of consecutive layers are achieved through the \xbar. In this mode, events from the \collector~can be redirected to any \slice, \textcolor{black}{output events are produced simultaneously to the input event processing, and all the layers of the network can execute in parallel.}
Alternatively, if the network needs to allocate more neurons than available in the \sne, intermediate feature maps (output events) must be stored in the external memory. In this case, the \sne~can be used in a time-multiplexed way to execute only a tile of the network. In this operating mode, synaptic connections are implemented by both the \xbar~and the \streamers~ through the external memory. 
\figurename \ref{fig:sample} shows the execution pipeline of operations to compute an eCNN layer. An input event is fetched and made available to all \clusters. Then, the \sne~updates all the neurons of each \cluster~that are sensitive to the current input event, \textcolor{black}{this operation is performed in 48 clock cycles}. The state of each output neuron is held across multiple input event processing, and as soon as a firing operation is received, all the neurons having the \arpan{membrane} potential above the threshold fire an output event.


%% file: content/simulation_results.tex
\section{Experimental results}
\label{sec:results}

This section provides post-synthesis estimates for the \sne~as a standalone engine. We synthesized the accelerator with \textit{Synopsys Design Compiler 2020.09}, in \textit{GlobalFoundries} 22nm FDX process. Specifically, we used 8T, 20, 24, 28, L, and SL voltage threshold cells, SSG corner, 0.72V nominal supply voltage, -40C, 400MHz target clock frequency.
Power consumption estimates have been performed at target 400MHz clock frequency, TT corner, 0.8V supply voltage, 25C, by using \textit{Synopsys Prime-Power 2019.12}. 

\subsection{SNE energy efficiency benchmark}
In this subsection, we evaluate how the performance, power consumption, area, and energy efficiency scale when the \sne~number of \slices~is configured to 1, 2, 4, and 8. We report the results for the \sne~configured with 16 \cluster~per slice and 64 TDM neurons per \cluster. 

\subsubsection{Area exploration}
\begin{figure}[t]
    \centering
    \centerline{\includegraphics[width=1\linewidth]{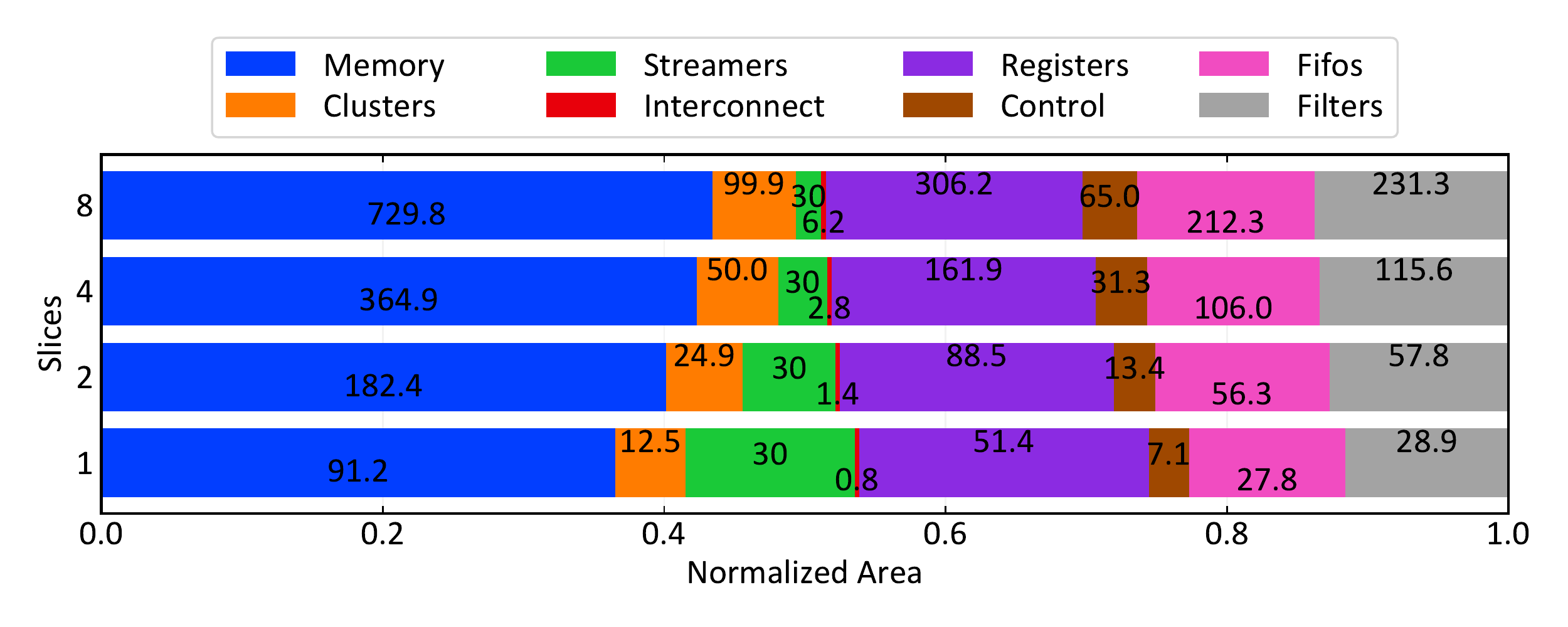}}
    \caption{SNE area breakdown for a different number of Slices. Values on the plot report the absolute area in kGE}
    \label{fig:plot_area}
\end{figure}

\arpan{
\figurename \ref{fig:plot_area} reports the area estimation in gate equivalent (kGE) for each configuration. This number is obtained as the total area estimate in \SI{}{\micro \meter \squared} provided by the synthesis tool, divided by the area of an ND2X1 gate (8T library).
Most of the area is occupied by latch-based memories holding the neuron state. As the number of \slices~increase, the \slices~and \xbar~area scales proportionally. \streamer~area remain constant. Area exploration shows that the fixed cost of the \streamers~is progressively absorbed by the data path area increase.
}

\subsubsection{Power analysis}

\begin{figure}[b]
     \centering
    \subfloat[Power consumption at average network firing activity of 5\%) \label{fig:power}]{\includegraphics[width=0.22\textwidth]{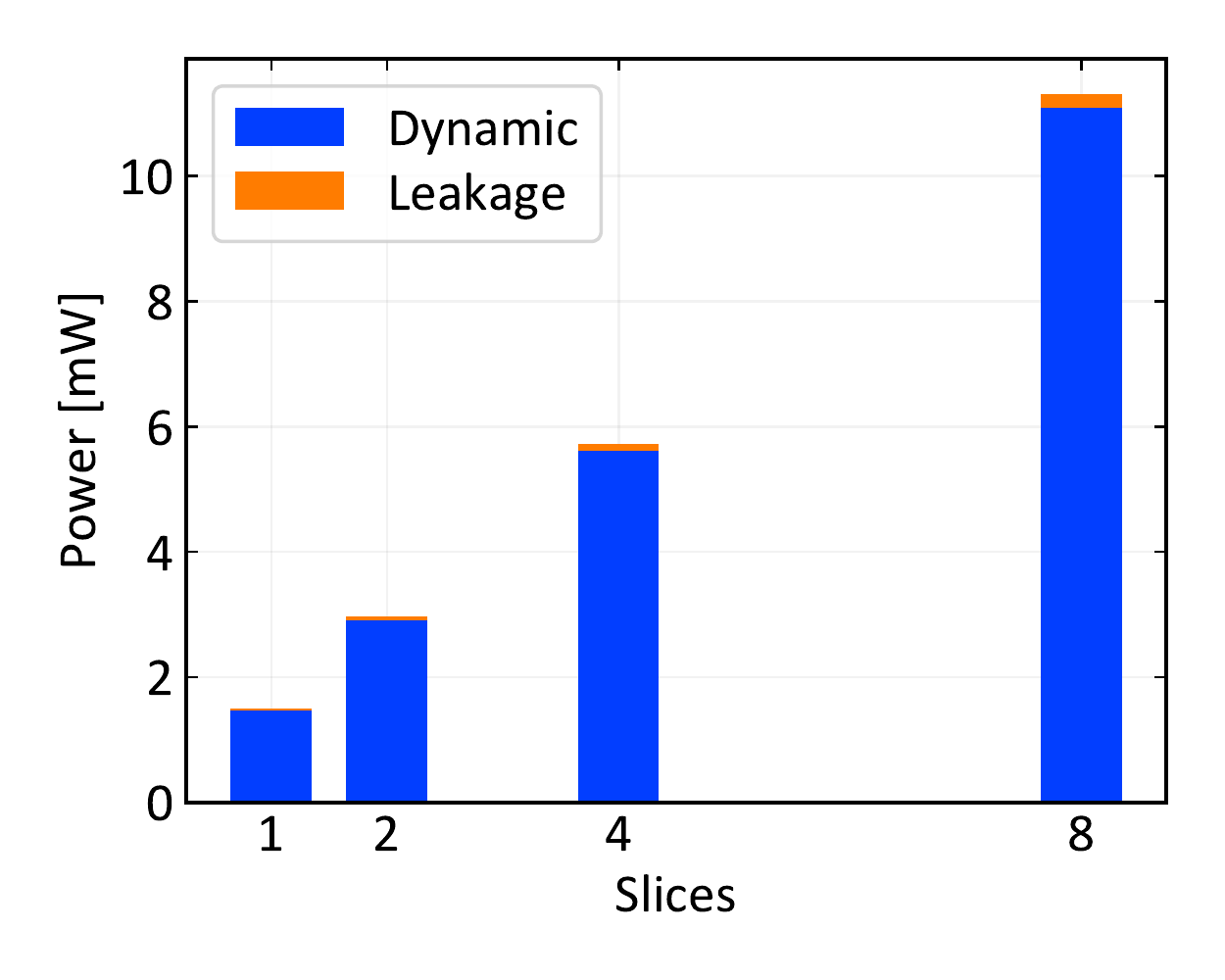}}
    ~
    \subfloat[Performance and energy per operation versus Number of Slices. \label{fig:performance}]{\includegraphics[width=0.22\textwidth]{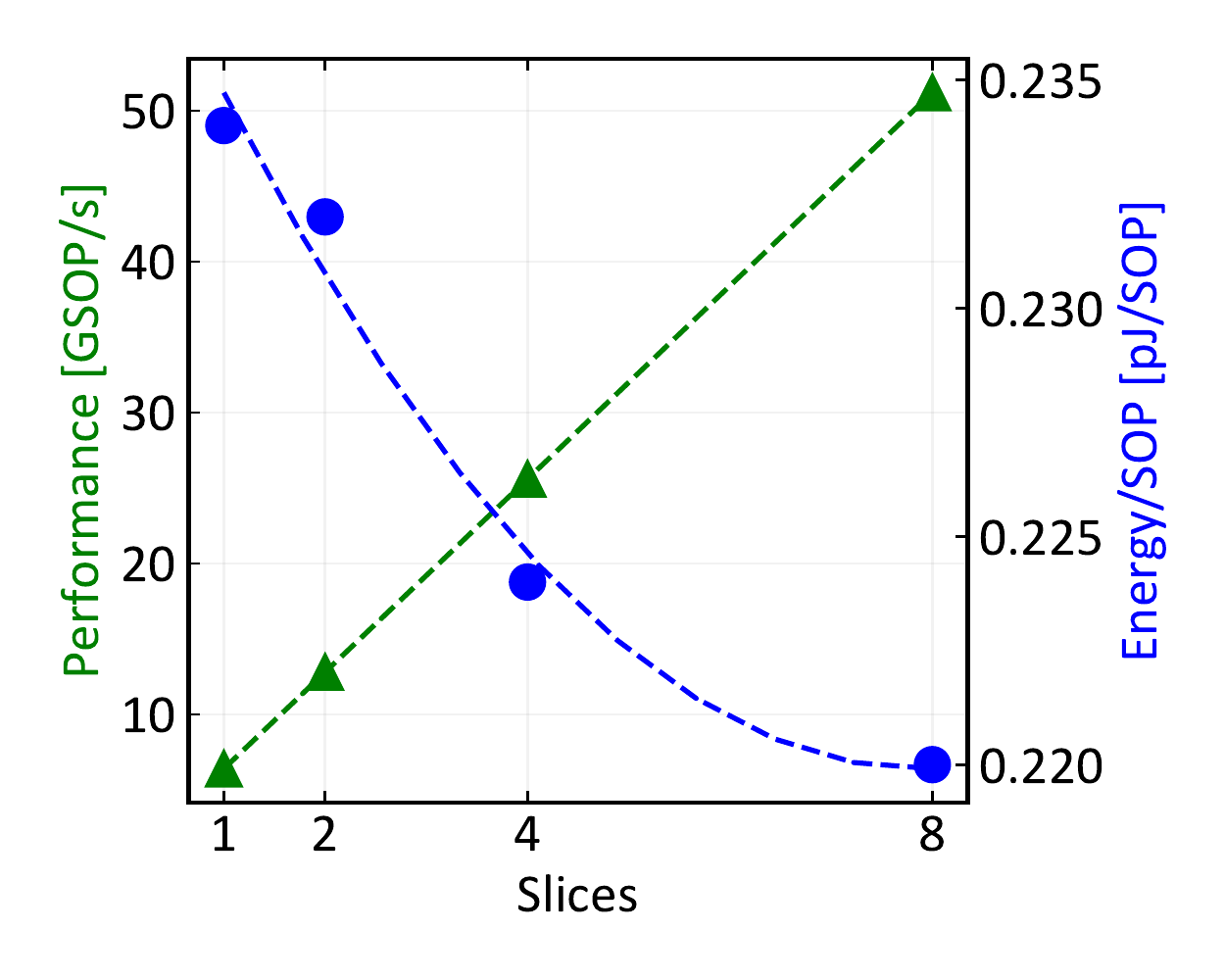}}
    \caption{\sne~energy and power consumption.}
\end{figure}

We estimated the power consumption using the simulated value change dump (VCD) activity of the post-synthesis netlist. The benchmark used for power consumption estimation is a sample eCNN layer where input events cause a neuron state update on all the \slices~and all \clusters~of each \slice. Input events are distributed across 100 time steps, and the layer is generating 5\% output event activity, which is comparable with the average network activity observed for the IBM-DVS Gesture data set and reported in section \ref{sec:benchmark}.
The VCD file used to extract the switching activities has been generated with Questasim-10.6b, while the power consumption has been estimated with Synopsys PrimePower-2019.12. \figurename \ref{fig:power} reports the power consumption for the different \sne~configurations. Dynamic power significantly dominates the total power consumption. Notice that the power consumption reported for this experiment is a worst-case estimate, as all computational units of the \sne~are updating the internal state of their neurons.

\begin{table}[t]
\centering
\resizebox{0.46\textwidth}{!}{%
\begin{tabular}{@{}lcccc@{}}
\toprule
\textbf{Data set} &
  \textbf{\begin{tabular}[c]{@{}c@{}}SNN \\ (SLAYER-SRM)\end{tabular}} &
  \textbf{\begin{tabular}[c]{@{}c@{}}eCNN \\ (SNE-LIF-4b)\end{tabular}} &
  \textbf{\begin{tabular}[c]{@{}c@{}}Inf. energy \\ {[}uJ/inf{]}\end{tabular}} &
  \textbf{\begin{tabular}[c]{@{}c@{}}Inf. rate \\ {[}inf/s{]}\end{tabular}} \\ \midrule
NMNIST &
  97.81\% &
  97.88\% &
  43 - 142 &
  261 - 79.5 \\
IBM DVS Gest. &
  92.42\% &
  92.80\% &
  80 - 261 &
  141 - 43 \\ \bottomrule
\end{tabular}%
}
\caption{eCNN classification accuracy, energy per inference and inference rate}
\label{tab:network_results}
\vspace{-10pt}
\end{table}

\subsubsection{Performance and energy efficiency benchmark}

\begin{table*}[t]
\centering
\resizebox{\textwidth}{!}{%
\begin{tabular}{@{}cccccccccccccc@{}}
\toprule
\textbf{Name} & \textbf{Tech.} & \textbf{\begin{tabular}[c]{@{}c@{}}Neuron \\ model\end{tabular}} & \textbf{Learning} & \textbf{Type} & \textbf{\begin{tabular}[c]{@{}c@{}}Neuron\\ number\end{tabular}} & \textbf{\begin{tabular}[c]{@{}c@{}}Neuron \\ area\\ {[}um2{]}\end{tabular}} & \textbf{\begin{tabular}[c]{@{}c@{}}Perf.\\ {[}GOP/s{]}\end{tabular}} & \textbf{\begin{tabular}[c]{@{}c@{}}Eff.\\ {[}TOP/s/W{]}\end{tabular}} & \textbf{\begin{tabular}[c]{@{}c@{}}Energy/SOP\\ {[}pJ{]}\end{tabular}} & \textbf{\begin{tabular}[c]{@{}c@{}}Freq.\\ {[}MHz{]}\end{tabular}} & \textbf{\begin{tabular}[c]{@{}c@{}}Power\\ {[}mW{]}\end{tabular}} & \textbf{bits} & \textbf{V} \\ \midrule
\textbf{\begin{tabular}[c]{@{}c@{}}SNE (this work)\end{tabular}} & \textbf{Digital 22nm} & \textbf{LIF} & \textbf{offline} & \textbf{\begin{tabular}[c]{@{}c@{}}Conv SNN\end{tabular}} & \textbf{8192} & \textbf{19.9} & \textbf{51.2} & \textbf{4.54} & \textbf{0.221} & \textbf{400} & \textbf{11.29} & \textbf{4} & \textbf{0.8} \\
Tianjic \cite{Pei2019} & Digital 28nm & - & - & Hybrid & 40000 & 361 & 649 & 1.28 & 6.18 & 300 & 950 & 8 & 0.9 \\
Dynapsel \cite{dynapsel} & Analog 28nm & - & online STDP & - & 256 & 150390 & - & 0.6 & 2 & - & - & 4 & 1 \\
ODIN \cite{frenkel_odin} & Digital 28nm & Bio Plaus. & - & - & 256 & 335.9 & 0.038 & 0.079 & 12.7 & 75 & 0.477 & - & 0.55 \\
TrueNorth \cite{truenorth} & Digital 28nm & EXP LIF & online & SNN & 1e6 & 389 & 58 & 0.046 & 27 & Asynch & 65 & 1 & 0.75 \\
SPOON \cite{Frenkel2020A2C} & Digital 28nm & - & DRTP & Conv SNN & - & - & - & - & 6.8 & 150 & - & 8 & 0.6 \\
Loihi \cite{loihi} & Digital 14nm & LIF+ & online  STDP & SNN & 131072 & 396.7 & - & - & 23 & Asynch & - & 1-64 & - \\
SpiNNaker 2 \cite{spinnaker2} & Digital 22nm & Prog. & - & DNN/SNN & - & - & - & 3.26 & 1700 & 200 & - & var. & 0.5 \\
\bottomrule
\end{tabular}%
}
\caption{State of the sne comparison. \\ 
\footnotesize 
}
\label{tab:comparison}
\end{table*}

\figurename \ref{fig:performance} shows the accelerator performance reported as synaptic operations per second (SOP/s). \sne~performance scales proportionally to the number of slices, as they operate independently, and the output bandwidth does not represent a bottleneck.
Note that in the case where more \slices~are added to the \sne, or when more activity is expected on the output of each \slice, the \sne~can be configured with a higher number of \streamers~to sustain the \slices~output bandwidth. \figurename \ref{fig:performance} also reports the energy per synaptic operation (SOP). This value has been calculated by dividing the energy consumed in a single cycle by the number of neuron updates performed in parallel. We remark that \sne~takes 48 clock cycles to consume an input event and update all membrane potentials serially. Therefore, true energy-to-information processing proportionality is ensured by design, i.e. the more events are present in a given input event stream, the more time \sne~spends to consume such an event stream.
We obtained the lowest energy/SOP when \sne~is configured for 8 \slices, consuming a constant energy of 0.221pJ/SOP. In this configuration, the ratio between the power consumed by the engines and other parts of the system is maximized, and most of the energy is spent on computation. 

\subsection{Accuracy benchmark}
\label{sec:benchmark}
\begin{figure}[t]
    \centering
    \centerline{\includegraphics[width=0.8\linewidth]{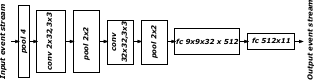}}
    \caption{SNN network topology used for the accuracy benchmark.}
    \label{fig:networks}
\end{figure}

We conducted two experiments to evaluate the accuracy of eCNNs deployed on the \sne. We trained the same network, whose topology is reported in figure \ref{fig:networks}, on two event-based data sets. The networks have been trained with a supervised approach. Specifically, we used back-propagation-based training in the SLAYER \cite{Shrestha2018} framework. 
As the \sne~implements a quantized variant of the LIF dynamics, where a linear decay has approximated the exponential decay, we implemented our \sne~neuron model and replaced the default \textit{SLAYER} spike response model (SRM) \cite{GERSTNER2001469}. We trained the same network with SRM neurons as a baseline comparison for our experiments. 

In the first experiment, we trained the network on the NMNIST data set\footnote{https://www.garrickorchard.com/datasets/n-mnist}, and we evaluated its accuracy. In the second experiment, we followed the same approach to train and evaluate the network accuracy on a more complex task, the IBM-DVS-Gesture data set\footnote{https://www.research.ibm.com/dvsgesture/}. We used 65\%, 10\%, and 25\% of samples for training, validation, and test set on the IBM-DVS Gesture data set, respectively. On the NMNIST data set, we divided the samples into 75\%, 10\%, and 15\% of samples for training, validation, and test set, respectively. On both data sets, \sne~eCNNs slightly improved the classification accuracy; accuracy results for both data sets are reported in Table \ref{tab:network_results}.

As a further investigation, we estimated the maximum activity of each eCNN layer, obtaining that a sample extracted by the IBM DVS-Gesture data set generated a firing activity between 1.2\% and 4.9\%, on average, across the entire network. \textcolor{black}{As \sne~consumes an input event in 48 clock cycles}, we used each layer activity to estimate then best case and worst case inference time, when the accelerator is clocked at 400MHz. \textcolor{black}{In this operating point, an input event is consumed in \SI{120}{\nano \second}, based on this value and the network activity, the inference is performed in a best and worst case time interval of 7.1ms and 23.12ms, respectively}. Similarly, we estimated that in this operating condition the \sne~can perform a new inference at a rate comprised between 141inf/s and 43inf/s, consuming a total inference energy between \SI{80}{\micro \joule}/inf and \SI{261}{\micro \joule}/inf, respectively.
Table \ref{tab:network_results} reports the \sne~inference performance and inference energy results for both data sets.

\subsection{Comparison with the state of the art}
In Table \ref{tab:comparison} we compare the \sne~to state-of-the-art neuromorphic engines implementing comparable neuron models and accelerating similar neural network topologies. To have a fair comparison with other engines reported in Table \ref{tab:comparison}, we remark that \sne~does not provide online learning capabilities. Compared to other architectures, \sne~shows both the lowest energy per operation 0.221pJ/SOP and the highest energy efficiency 4.54TSOP/s/W, while reaching SoA accuracy on event-based data sets; 92.8\% on IBM DVS-Gesture. This result narrows the gap between neuromorphic platforms and classical DNN accelerators\cite{classical_dnn}. The energy efficiency reported on \sne~improves by 3.55X the energy efficiency reported by Pei et al. \cite{Pei2019}. \textcolor{black}{Note that area-wise, both designs are implemented in a comparably scaled technology node. Additionally, assuming the same 400MHz target frequency and extrapolating our results to the 0.9V operating condition, \sne~would still achieve 4.03TOP/s/W and consume 0.248pJ/SOP.}

%% file: content/conclusion.tex
\section{Conclusion}
\label{sec:conclusion}

In this paper, we presented a configurable digital engine for brain-inspired event-based convolutional neural networks (eCNN). Our accelerator exploits the unstructured sparsity of data produced by event-based sensors by performing a number of operations proportional to the input stream activity. Our engine consumes explicitly spatial and temporal-encoded input events to achieve high energy efficiency, selectively updating the internal output neurons states. \textcolor{black}{We demonstrated that \sne~can reach SoA 98.2\% classification accuracy on the IBM DVS-Gesture data set while performing up to 141inf/s. \sne~achieves SoA energy efficiency of 4.54TSOP/s, comparable to classical DNN inference engines. Ultimately, \sne~consumes 0.221pJ/SOP, which is the lowest energy per operation reported on a neuromorphic platform to the best of our knowledge.}